# Hydrogel Leclanché Cell: Construction and Characterization

*G. Jenson[1,*], G.Singh[2], A. Ratner[2], J. Bhama[1†]*

[1] University of Iowa, Department of Surgery, Iowa City, Iowa, 52242
[2] University of Iowa, Department of Mechanical Engineering, Iowa City, Iowa, 52242
[†]Bhama-Ratner Artificial Heart and MCS Advancement Lab (BRAHMA) Lab, Department of Surgery, Carver College of Medicine, University of Iowa, Iowa City, Iowa, 52242
[*]E-mail: greg1.jenson1@gmail.com

A liquid-to-gel based Leclanché cell has been designed, constructed, and characterized for use in implantable medical devices and other applications where battery access is limited. This well-established chemistry will provide reliable electrochemical potential over a wide range of applications and the novel construction provides a solution for the re-charging of electrodes in hard to access areas such as an internal pacemaker. The traditional Leclanché cell comprised of zinc (anode) and manganese dioxide (cathode), conductive carbon powder (acetylene black or graphite), and aqueous electrolyte hydrogel ($NH_4Cl$ and $ZnCl_2$) has been suspended in an agar hydrogel to simplify construction while maintaining electrochemical performance. Agar hydrogel, saturated with electrolyte, serves as the cell support and separator allowing for the discharged battery suspension to be easily replaced once exhausted.

## 1.INTRODUCTION

The first clinical cardiac pacemaker was implanted in 1958 by Ake Senning. However, the electrochemical power supply only lasted a few hours. This groundbreaking advancement inspired collaborative effort to overcome the challenge to produce a reliable power supply for medical implants [1]. When considering the design of batteries for implantable medical devices, additional challenges arise when choosing chemical components, electric cell design, and most importantly how to recharge the electrochemical cell. Today, various lithium ion technologies have significantly extended the battery lifetime however, replacing the battery requires a surgical procedure that may result in further complications [2, 3]. We propose an alternative approach to recharge the battery through the re-development of Leclanché electrodes. These battery components may be injected through a needle, simplifying the medical procedure and decreasing the risk to the patient.

In pursuit of this idea we have used the Leclanché type chemistry to demonstrate the proof of concept of our liquid battery because this well characterized chemistry boasts high energy density and reliability [4, 5]. In addition to favorable electrical characteristics, we have also evaluated the current state of battery and fuel cell technology. We believe that repurposing established battery technology exemplifies creativity, broadens the landscape for electrochemical possibilities, and serves to establish new methods to further electro-mobility [6-8]. This well-established chemistry has served as a reliable power source for more than a century and high performance commercial batteries have been utilized in a variety of applications since the 1970's [9]. Recent studies regarding electrolyte composition, conductive additives, and new applications, such as flow batteries, demonstrate the relevance and applicability of the original Leclanché cell [10-12].

The first zinc carbon primary is known as the Leclanché cell and was developed over a century ago by a French telegraph engineer, Georges Leclanché [4]. Originally, these primary cells were formulated with zinc anode and manganese oxide cathode in combination with various conductive carbon powders saturated with an aqueous electrolyte. Later, Carl Gassner reconstructed these batteries with powdered or paste electrodes to develop the first "dry cell" battery. Dry cells are ideal for mobile applications that require intermittent discharge, such as flashlights or radios, where leakage of liquid cells is detrimental. Today, the commercial Leclanché is housed in a cylindrical zinc container that encapsulates the separator, manganese based cathode and a carbon rod current collector. Often the mobile electrolytes, ammonium chloride and zinc chloride, are suspended in a separator such as: porcelain, paper, or gelling agents such as starch. Modest cost, simplicity of production, and favorable discharge characteristics have made this chemistry desirable for a variety of applications and has proven to occupy a significant portion of the market [9,13].

In this research, the original Leclanché chemistry and battery construction has been simplified to be used in internal medical devices. The electrodes have been suspended in an agar hydrogel to simplify the construction and replacement/recharging procedure of spent batteries. While under electrochemical load the hydrogel provides mechanical support allowing the battery to sample many spatial confirmations to maintain energy density. Hydrogels are characterized by polymeric networks than can absorb a significant amount of water and become swollen granting new physical properties, such as physical stability and ionic conductivity [14]. Hydrogels as energy storage materials have seen recent interest and have been used in a variety of electronic devices such as capacitors, sensors, and scaffolds for catalysis [15-18]. Excitingly, there has been some research and characterization of the zinc $MnO_2$ alkaline cell in gel form, but the application and reproducibility has yet to be demonstrated [19]. Additionally, our hydrogel battery features a simple construction due to the fact that the hydrogel also serves as the separator allowing the entire battery hydrogel to be constructed in a simple container. Gel based separators are desirable because of high ionic conductivity, ability to form a variety of shapes, and ease of fabrication [20]. Various carbohydrates binders and electro-spun modified agar have been reported for use as a battery separator however, we are not aware of literature reports using unmodified agar as a battery separator [21]. Due the instability of agar, we demonstrate the recharging of these electrochemical cells by treating the exhausted agar electrode matrix with warm aqueous acid to remove and replace the cell with new active material [22].

## 2. EXPERIMENTAL

*2.1 Design of Experiment*

The experimental setup has been designed to calculate the discharge rate of a given battery. The system can acquire analog data from a maximum of 16 data inputs in parallel and log it appropriately in a comma-separated values (.*CSV*) format (Figure 1).

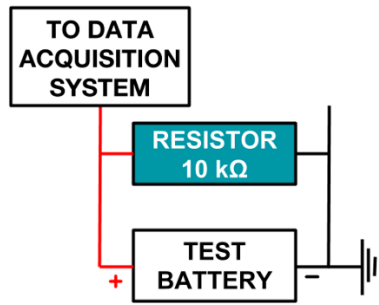

**Figure 1**: A single-battery test setup

To calculate the discharge rate, a given battery is connected to a 10 kΩ resistor and is discharged across it until empty. Voltage across the resistor is measured using an acquisition card (more details in next section) at 30-minute intervals. Plotting the voltage values of the battery at different points of time shows the required discharge trend line and helps compare different batteries.

Once set, the experiment can simultaneously log data from a variety of batteries for weeks on end without intervention or supervision. The .*CSV* file generated by the system can then be used in a standard program like MS Excel or MATLAB to generate data plots.

### 2.2 Data Acquisition System

The data acquisition system consists of an acquisition card and a data logger. The acquisition card used here is an Arduino ATMega 2560, which is an electronic microcontroller that can be programmed to acquire analog data from a maximum of 16 channels at any given interval. Signal range is from 0-5 V, and resolution is 4.9 mV [23].

The acquired signal is sent via USB to a data logger, which is a Raspberry Pi 3 Model B running Raspbian. It is a Linux-based, single-board mini-computer. It can provide a GUI interface output to a computer screen via HDMI, and peripherals like a keyboard and mouse can be plugged into it (Figure 2).

The data logger runs a Linux-based version of Arduino. The acquisition card conditions the signal into a series of comma-separated values, which can be logged on to the Arduino Serial Monitor screen. These can be saved directly as a .*CSV* file.

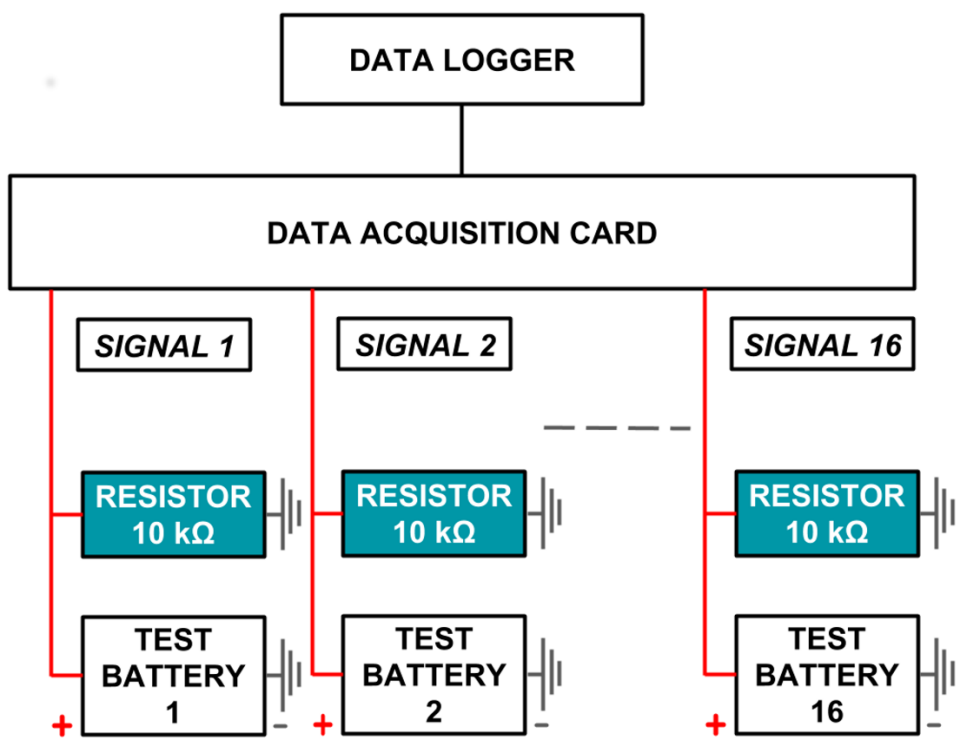

**Figure 2**: Schematic of data acquisition system

*2.3 Gel Leclanché Cell Construction*

An Agar-Leclanché-graphite cell is constructed in a 20 ml vial by mass or volume of components. The electrolyte hydrogel is used at various points in the method and consists of the following by weight: 26% $NH_4Cl$ (RPI, 99.5%), 8.8% $ZnCl_2$ (Sigma, 98%), and 65.2% DI $H_2O$. Anode hydrogel is constructed by heating 4 g Zn powder (Sigma, 10 **μ**m, 98%), 2 g graphite powder (Sigma, 20 **μ**m), and 6 ml electrolyte hydrogel until dissolved. The hydrogel is vortex mixed, and 3 g of the anode hydrogel is transferred to a new vial with 3 g of a warm 1% w/v agar (RPI) electrolyte hydrogel. The warm agar separator hydrogel (3 g) was layered on the anode hydrogel followed by 3 g of cathode hydrogel, which consists of $MnO_2$ 8 g (Sigma, 99%), graphite 4 g, and 15 ml electrolyte hydrogel. 5 mm carbon electrodes (Eisco) are coated in paraffin (Gulf Wax) to span the cathode hydrogel and are submerged in the fuel cell for characterization (Figure 3).

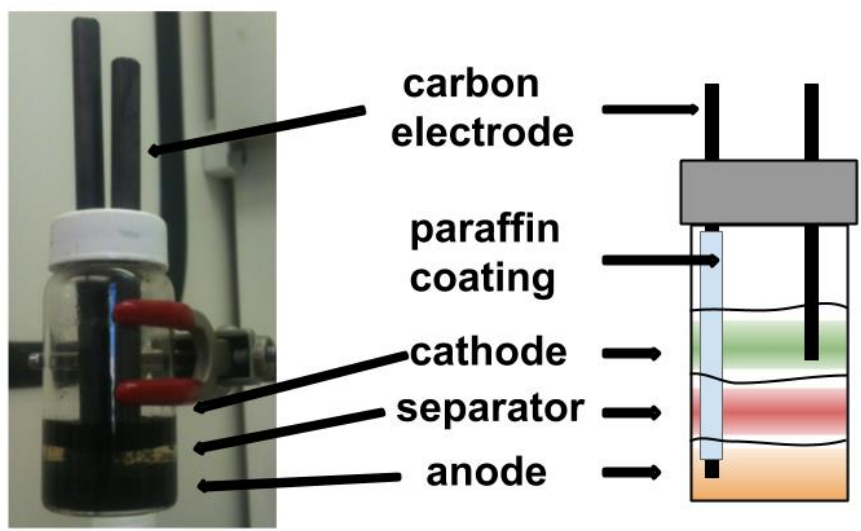

**Figure 3**: Gel Leclanché-graphite cell constructed in a 20 ml vial.

*2.4 Gel Leclanché Cells with Various Conductive Additives*

After the initial characteristics and experimental setup of the Zn agar cells were defined, the quantities of Zn and conductive additive were varied using a similar cell design. The anode hydrogel consists of 2 g saturated conductive additive electrolyte hydrogel (1.5 g acetylene black or graphite and 20 ml electrolyte hydrogel), 1 g agar electrolyte hydrogel (2.5% w/v) and 0.25 g, 0.5 g, or 0.75 g Zn powder. Warm agar electrolyte separator (3 g), as discussed previously, is layered on the anode and allowed to cool. The warm cathode hydrogel is layered on the separator and consists of 2 g of the following mixture: $MnO_2$ 3 g, 6 g acetylene black electrolyte hydrogel, and 3 g 2.5% w/v agar electrolyte hydrogel. 0.5 mm carbon electrodes coated in paraffin to span the cathode are used as current collectors (Figure 3).

*2.5 Closed Leclanché Gel Cell Construction and Purging*

The gel Leclanché cell is constructed in an 11 ml glass chamber equipped with purge/fill ports and current collection ports (Figure 4). The gel anode and cathode consist of 0.5% w/v agar dissolved in 3 ml aqueous electrolyte and ethylene glycol hydrogel (Sigma) (1:1) with 0.1 g acetylene black and 0.25 g $MnO_2$ or Zn. The gel separator consists of 0.25% w/v agar dissolved in aqueous electrolyte and ethylene glycol hydrogel (1:1). The cathode hydrogel was layered first, followed by the separator and, finally, the anode hydrogel. The original gel hydrogel was removed by treating the cell with warm acidic water (pH ~1) until clean, and the next Leclanché gel components were layered as previously described.

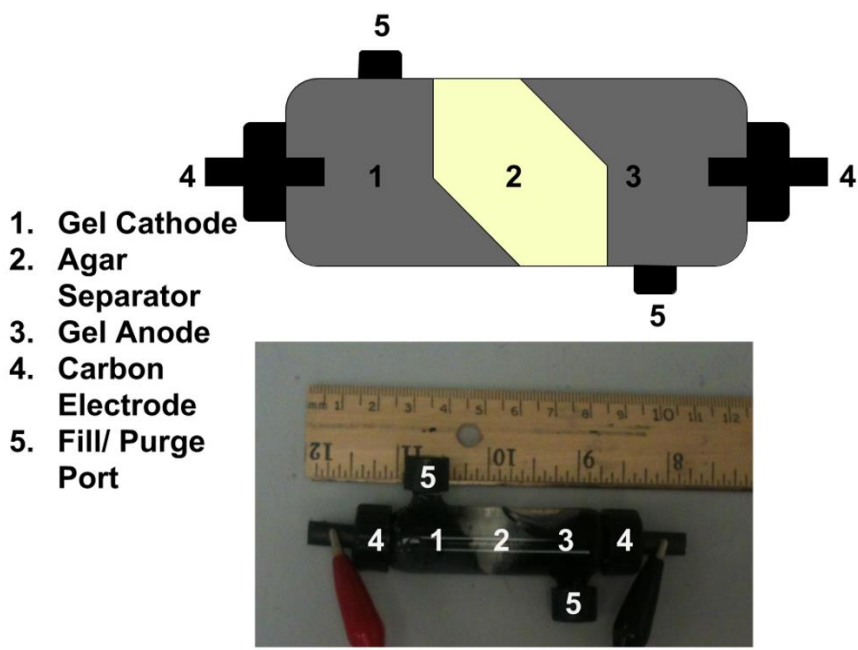

**Figure 4**: Closed gel Leclanché cell construction in an 11 ml glass chamber with ports for purge/filling and ports for current collection.

## 3. RESULTS AND DISCUSSION

### 3.1 Gel Leclanché Cell: Proof of Concept

Initially, the standard Leclanché chemistry was suspended in an aqueous agar hydrogel to explore the feasibility of the cell design and characterize the overall cell electrochemistry. Discharge characteristics of three trials are shown in Figure 5. Each cell was characterized by recording the voltage under a 10 kΩ load every 30 min until the cell potential reached a constant reading (~0.1V) . This resistor was chosen because it allowed the experiments to be conducted on a reasonable timeline that yielded data which compares some of the important battery components. Many battery standards use resistors in the ohm range, however in the interest of time our battery needed a larger resistor to generate complete discharge plots for each varied component [24].

The average open circuit voltage (*OCV*) of three electrochemical cells on construction (~1.4V) was larger than initial closed circuit voltage (*CCV*) (~0.75V) which is due to the electrochemical load imposed by the resistor. Typically the *CCV* approaches the *OCV* and can be optimized based on the electrolyte, cell construction, and separator which will be discussed later. The general discharge characteristics displayed by the Zn-graphite cells are similar to previous reports in the literature, however the timescales may differ depending on the load tested [25-26]. There is an initial rapid drop in potential (0-50000 seconds) followed by a period of stabilization (50000-240000 seconds) until the final rapid discharge (**Figure 5**).

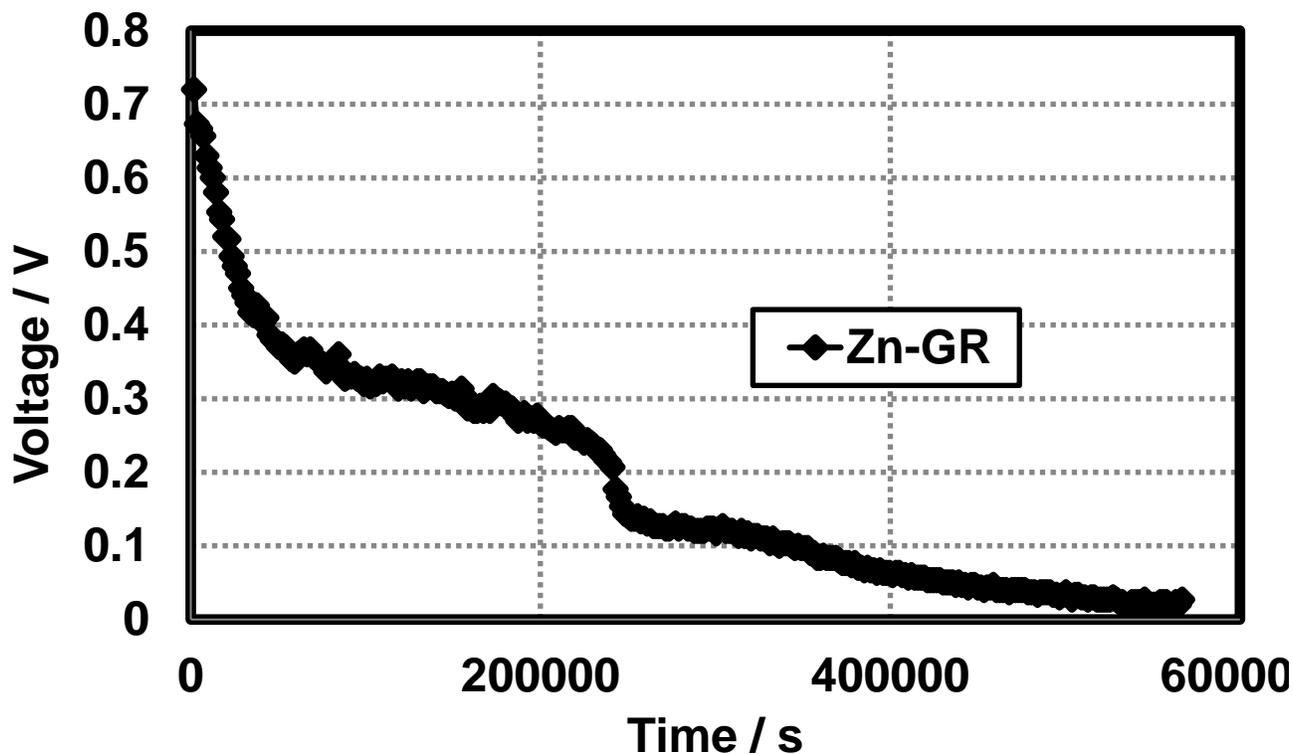

**Figure 5**: Gel Leclanché cell continuous discharge data across a 10 kΩ resistor with GR as the conductive additive. Average of three trials.

*3.2 Gel Leclanché Cell: Different Conductive Additives*

Zinc agar batteries constructed with acetylene black as the conducting additive display slightly enhanced characteristics under continuous discharge. Initial *OCV*'s, irrespective of the zinc quantity, in each cell is greater in cells constructed with acetylene black than cells constructed with graphite (~1.15V *vs*. ~0.75V). Additionally, the acetylene black cell produces a nearly constant *CCV* (~1.05V) from time zero until about 500000 seconds in cells constructed with 0.5g and 0.75g zinc powder. Beyond 0.5g zinc powder there is not a significant difference in the cell lifetime however, increasing the $MnO_2$ in combination with increasing zinc may deliver longer cell lifetimes. It has been previously reported that the specific sources of $MnO_2$ can have a profound effect on the cell performance which will be further optimized [26-7]

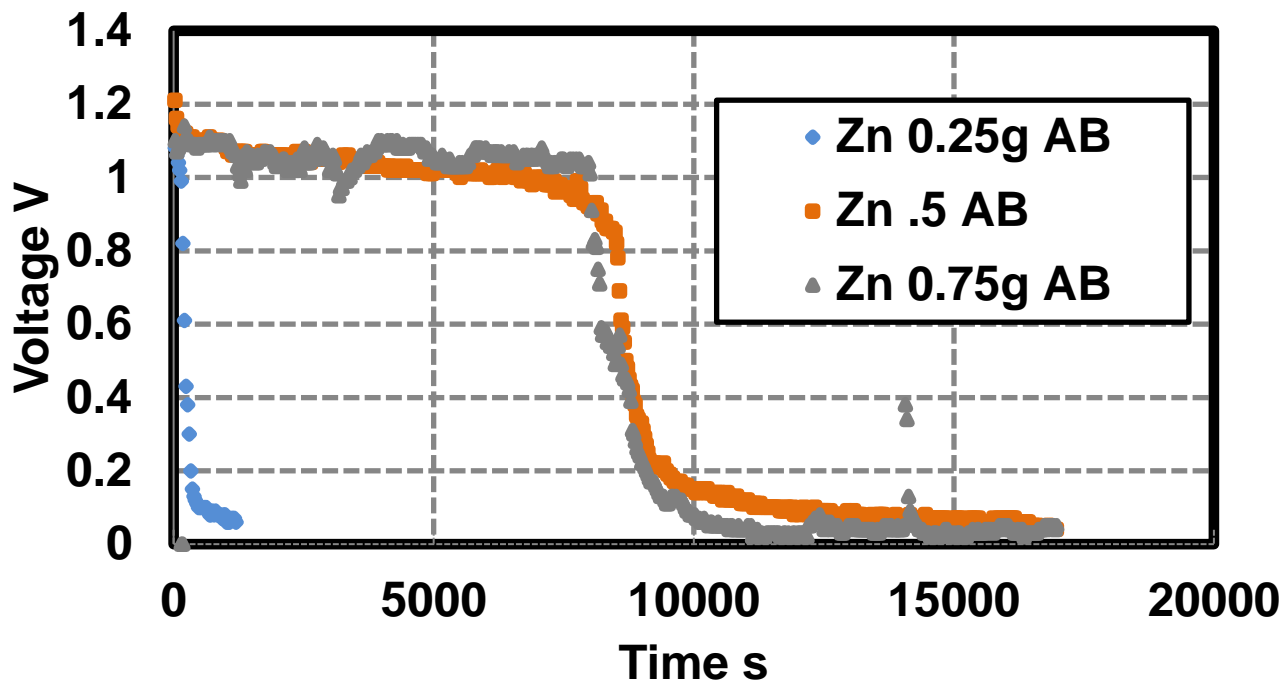

**Figure 6**: Gel Leclanché cell continuous discharge data across a 10 kΩ resistor with *AB* as the conductive additive, 0.5 g MnO$_2$, and variable zinc 0.25 g (◆) 0.5 g (■) 0.75 g (▲).

Variable Zinc-GR cells are constructed in a similar manner to the variable Zn-AB cells and display similar discharge characteristics to the initial Zinc-*GR* primary cells. On construction *OCV's* are dependent on the quantity of Zinc. Above 0.5g Zn powder, it appears that the *OCV* is not affected when the MnO$_2$ is held constant (0.5 g) and approach ~1.2V (Figure 7). Discharge characteristics of *GR* cells do not appear to rely heavily on the quantity of Zn and display a more rapid discharge rate when compared to *AB* cells. The terminal voltage appears to increase with the amount of Zn used in construction and is approached more gradually when compared to the *AB* cells. It must also be noted that the potential is more variable in *GR* based cells than the *AB* based cells.

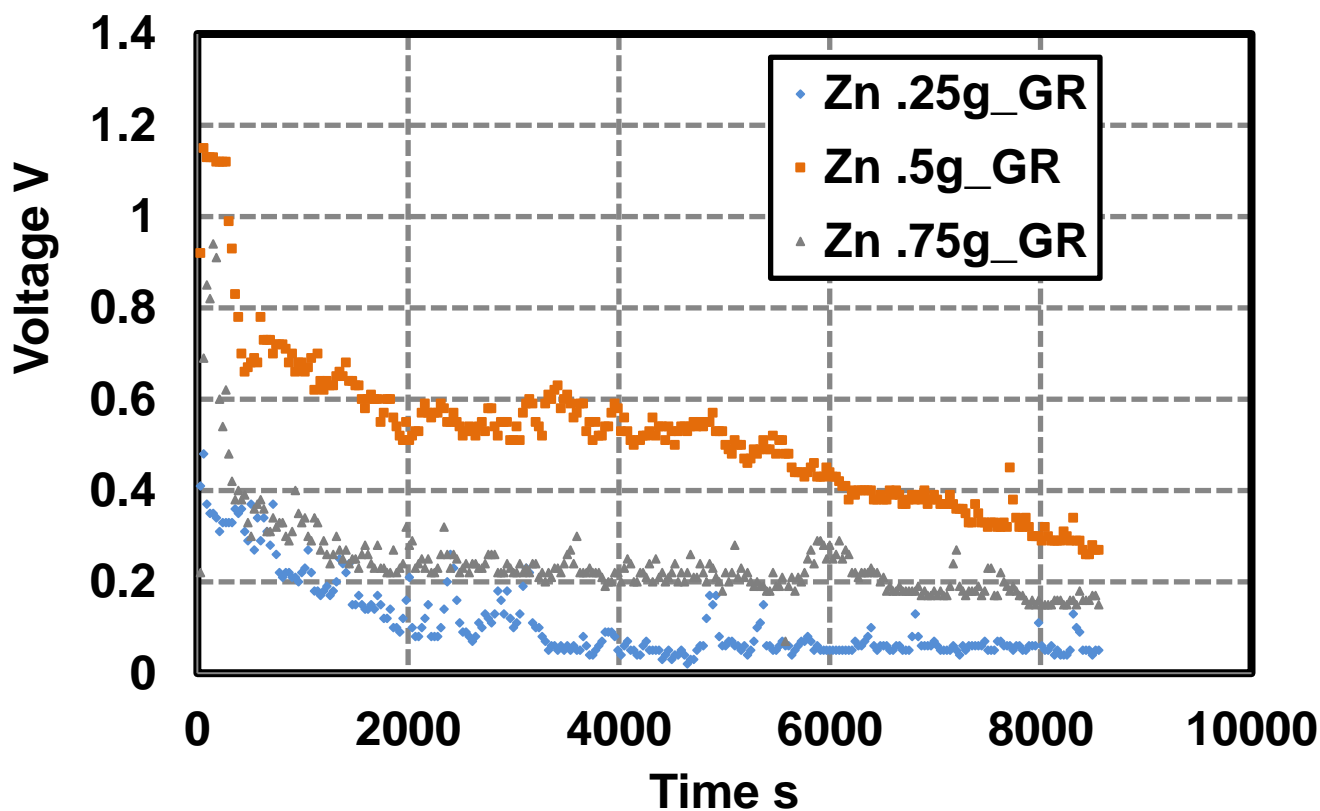

**Figure 7**: Gel Leclanché cell continuous discharge data across a 10 kΩ resistor with *GR* as the conductive additive, 0.5 g MnO₂, and variable zinc 0.25 g (🔷) 0.5 g (🟧) 0.75 g (🔺).

*AB* based cells display greater voltage stability and maintains higher voltage for prolonged time. This suggests that *AB* cells may better serve situations where higher voltage is required over greater time spans, while *GR* cells are better suited for rapid discharge over shorter time spans. It has previously been reported that the conductive additive can be used to tune Leclanché type cell discharge for specific applications and our data showcases the reproducibility of the agar based system [28-29].

The conductive additive also influences the energy density (JL^-1, total mAh, and the coulombic capacity (Ah) of cells constructed with different conductive additives. Although the quantity of active material does not change and therefore, theoretically the energy density should remain the same regardless of the conductive additive. We observed differences in the energy density as a function of the conductive additive and quantity of zinc (Figure 8). As expected, the energy density increased with increasing zinc.

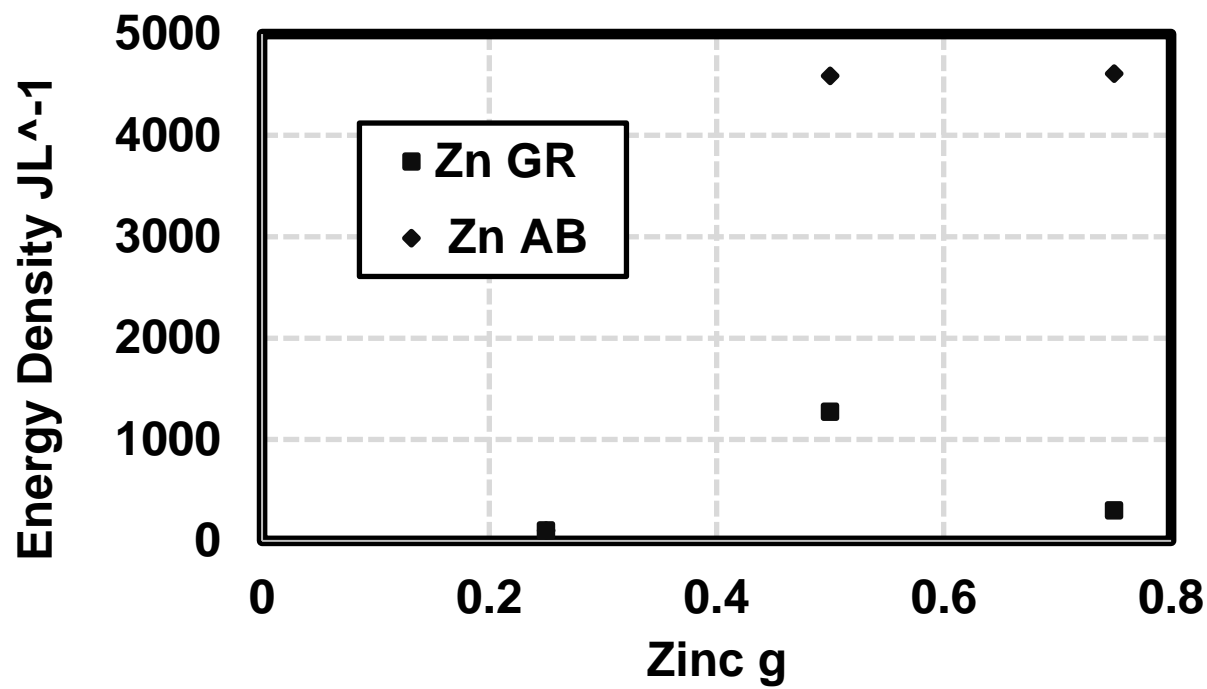

**Figure 8**. Energy density of cells made with variable amounts of zinc and different conductive additives (AB and GR)

However, the conductive additive also appears to influence the total mAh (Figure 9). At 0.25g zinc cells with AB and GR have similar energy densities (AB 102.14 JL^-1, GR 101.25 JL^-1) (Figure 9). Beyond 0.5 g zinc the difference in energy densities between cells made with AB and GR is large and appear to be great in cells made with AB.

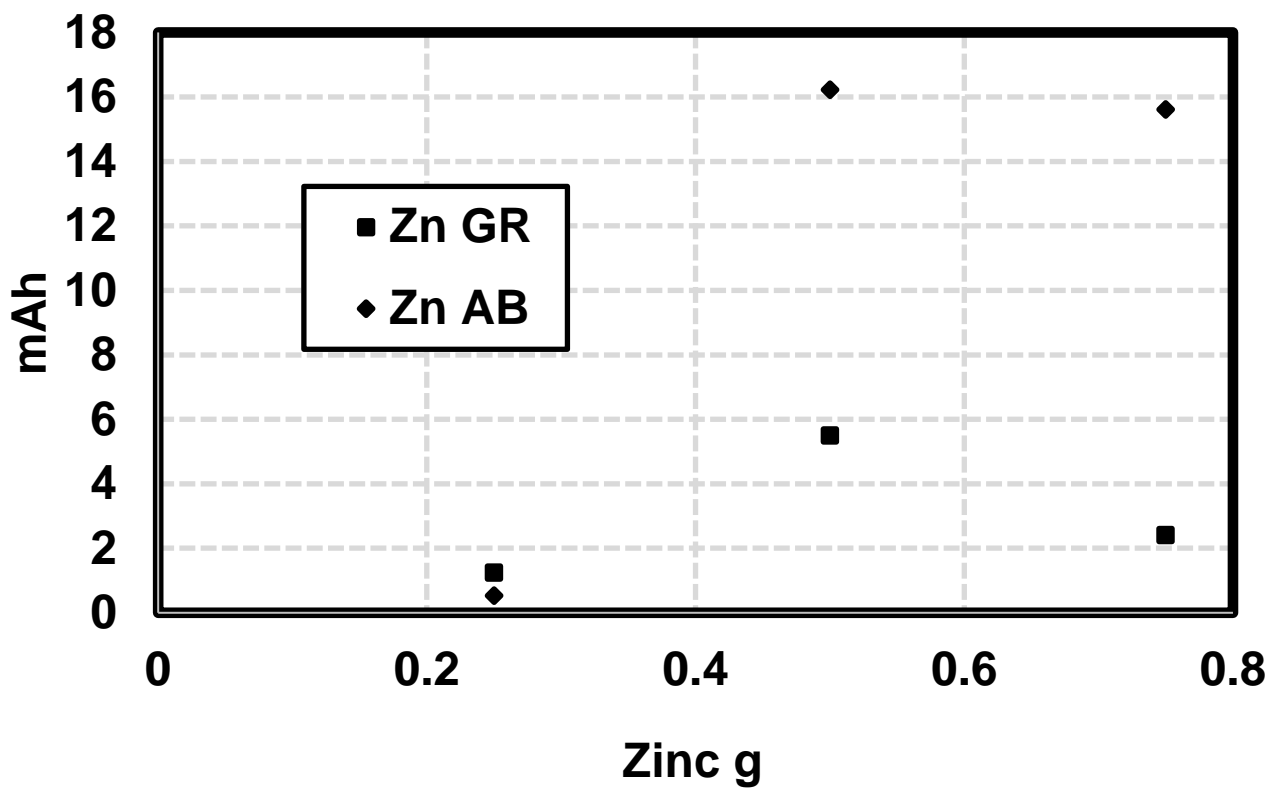

**Figure 9**. Total milliamp hours of cells made with variable amounts of zinc and different conductive additives (AB and GR)

  With regards to mAh, cells constructed with variable quantities of zinc and either AB or GR a similar trend to energy density is observed, greater zinc increases mAh with AB performing better than GR at larger quantities of zinc (Figure 9). It must be noted that the total mAh of cells made with 0.25g zinc and GR (1.24 mAh) is larger than cells made with AB (0.52 mAh). Further studies should be done to determine the validity of this finding however, it may suggest that, for applications where the size of the battery must be extremely small, GR should be used as the conductive additive.

  Total Ah were calculated for theoretical cells and compared to cells made with AB and GR at different quantities of Zn (Figure 10). Coulombic capacity is a way to relate the potential electrochemical energy of a material to the harvestable electrochemical energy of the system as reported in reference 9. Here we show that the constructed cells are able to perform near 1% of their theoretical coulombic efficiency. It is well known that primary cells do not reach the theoretical coulombic efficiency, but can be improved upon with different conductive additives, cell design, and separators.

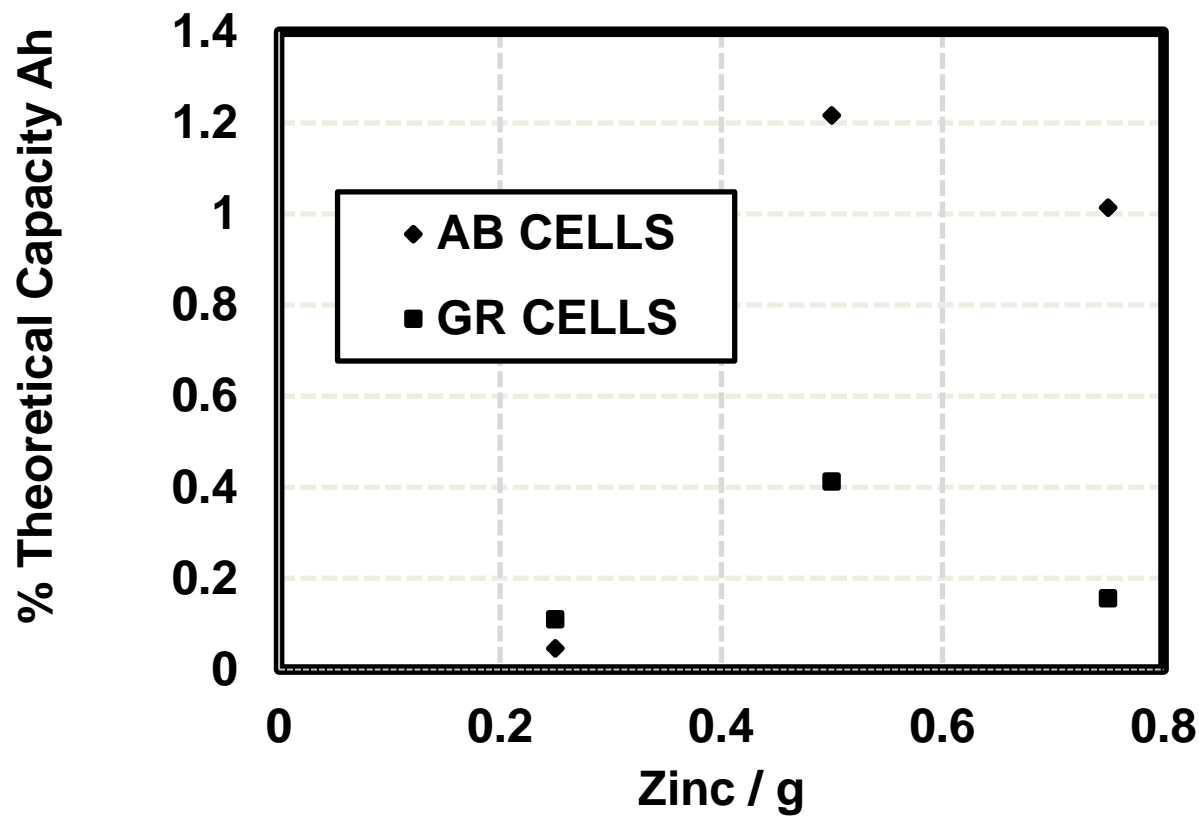

**Figure 10**. Coulombic efficiency as a % of theoretical coulombic capacity.

The ideal conductive additive should simply serve as a medium for electron transportation without influencing the chemical reactions of the cell. There are obvious trends in the OCV, CCV, discharge rates, energy density, and mAh depending on what conductive additive is used. There is considerable literature reporting physical characteristics of conducting carbon powders when used in their compressed form and limited literature regarding their use in other specialized applications [30,31]. It is generally agreed that the particle size and size range, morphology, source, and pressure used to compress powder influence the conductivity of the material and in general, smaller and more tightly packed particles conduct electrons better [32-34]. The GR particles used were <20µm in diameter and the AB particles were 25-45nm in diameter and it that the considerably smaller size of AB particles is responsible for the greater electrical performance of cells. Further studies cannot discount the influence of other physical characteristics such as adsorption, solubility, and chemical structure of the conductive additive to influence the electrochemical performance of the cell.

When comparing our battery to commercially available batteries and other reports in the literature we find that the OCV general discharge trends are similar and that notable differences are obvious. Differences displayed by our batteries include high internal resistance (Mohm range) as compared to other systems that boast ohm range. We expect to optimize our hydrogel as to decrease internal resistance and increase battery performance. Our energy density, and coulombic capacity is about (0.5-1%) of other reports, however we have not saturated our hydrogel with active material and believe this will be improved upon in the near future [35,36]. Our batteries, which contain the same active chemicals as other zinc-$MnO_2$ systems, are performing similar chemical reactions to those already established, however in a very different construction. We can be confident that our novel construction has not significantly altered the feasibility of this chemistry.

### 3.3 Gel Leclanché Cell Purge and Refill

To demonstrate the applicability of this method a closed glass cell was constructed to record fill and purge voltage of gel Leclanché cell. Ethylene glycol was added to make the purge and refill process easier and the influence on electrochemical performance in under investigation. Discharge data for gel based Leclanché cell under 10kΩ continuous load (Figure 11). Upon construction a potential of 0.15V is observed, a peak potential observed at 0.35V after 125000 seconds, and a stabilized potential of 0.29V is seen from 150000 seconds to 350000 seconds. At 400000 seconds the cell was purged and refilled and the potential again peaked a 0.28V.

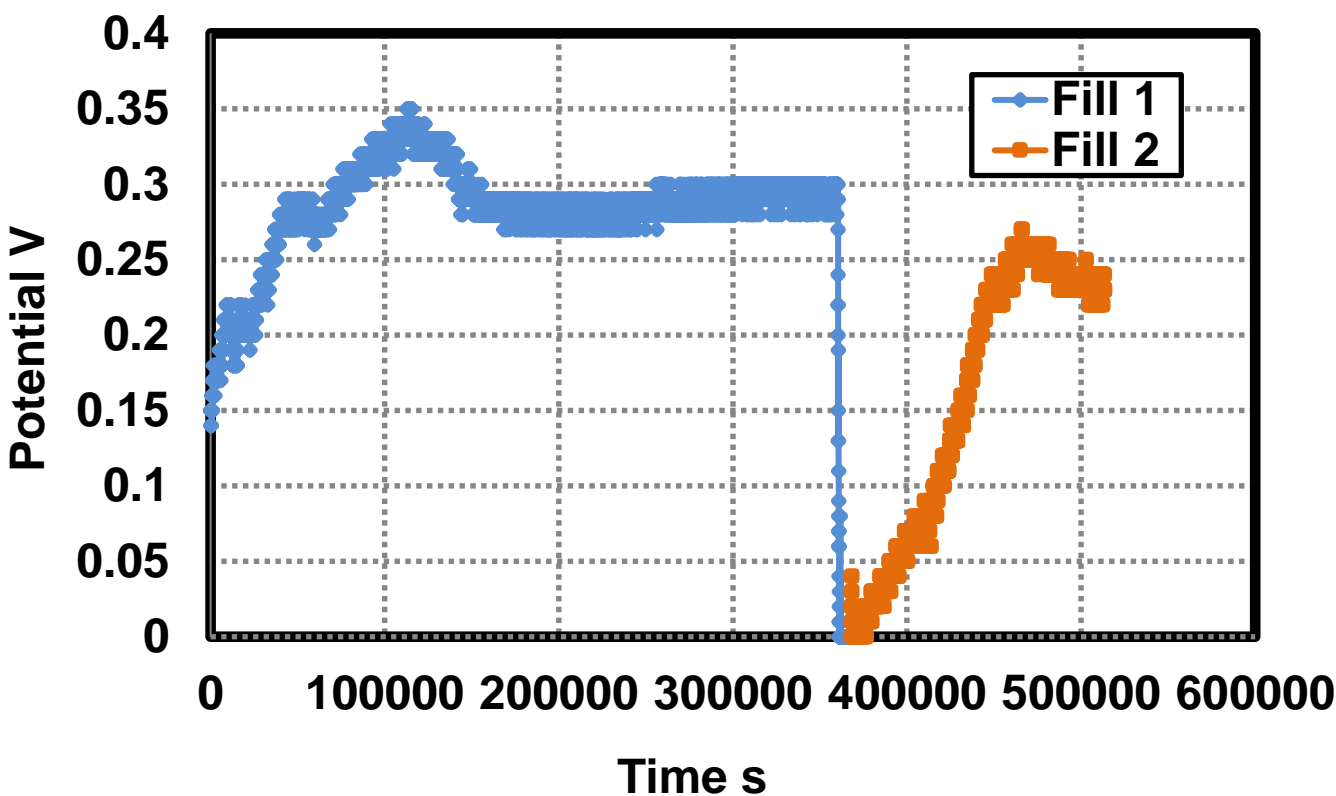

**Figure 11**.  Fill and purge voltage of a closed gel Leclanché cell under 10 kΩ load with *AB* conductive additive. The cell was purged and refilled at 400000 seconds.

## 4. CONCLUSIONS

The traditional chemistry of the Leclanché cell has been re-constructed in aqueous hydrogel for use in implantable medical devices or other hard access areas. In this work the standard Leclanché chemistry has been suspended in agar hydrogel to immobilize the anode and cathode as to prevent spilling in addition to serving as the battery separator. These cells are easily constructed and utilize components (carbon electrode, glass vials, agar separators) that are inexpensive and common in many research laboratories. Construction of these fuel cells in a single vial exemplifies the feasibility of this method for hard to access applications, such as a pacemaker. These implantable cells will be tuned to increase cell lifetime, voltage, and discharge characteristics that meet the energetic demands of a specific device. The role of the conductive additive plays a significant role in cell potential in addition to the discharge characteristics and will be exploited to suit particular energy demands. Fine tuning the removal and reconstruction of the spent agar electrode hydrogel is being refined. This preliminary data highlights the potential of this methodology as a technically simplify the procedure to replace exhausted electroactive material in devices where access is limited.

ACKNOWLEDGMENTS
The authors would like to acknowledge Dr. Ronald Weigel, Head of Department of Surgery and Professor of Biochemistry, for his support in securing lab space and providing financial assistance. This support has been instrumental in making this research a reality.
This research is funded, in part, by the Mid-America Transportation Center via a grant from the U.S. Department of Transportation's University Transportation Centers Program, and this support is gratefully acknowledged. The contents reflect the views of the authors, who are responsible for the facts and the accuracy of the information presented herein, and are not necessarily representative of the sponsoring agencies.

**References**
1. W. Greatbatch, *J.Cardiovasc. Nurs.,* 5 (1991) 80.
2. V. Sarma Mallela, V. Ilankurnaran, N. Srinivasa Rao, *Indian Pacing and Electrophysiol. J.,* 4 (2004) 201.
3. M. Kotsakou, I. Kioumis, G. Lazaridis, G. Pitsiou, S. Lampaki, A. Papaiwannou, A. Karavergou, K. Tsakiridis, N. Katsikogiannis, I. Karapantzos, C. Karapantzou, S. Baka, I. Mpoukovinas, V. Karavasilis, A. Rapti, G. Trakada, A. Zissimopoulos, K. Zarogoulidis, P. Zaro. Goulidis, *Ann. Transl. Med.*, 3 (2015) 45.
4. G.Leclanché, *Notes sur l'emploi des piles électriques en télégraphie, pile constante au peroxyde de manganèse à un seul liquide*, Impr. de Hennuyer et fils, (1867) Paris, France
5. J. Larcin, *Ph.D.Thesis*: *Chemical and electrochemical studies of Leclanché cells*, (1991) Middlesex University, London, United Kingdom.
6. U. Stimming, P. Bele, *J. Fuel Cells,* 16 (2016) 2.
7. U. Stimming, P. Bele, *J. Fuel Cells,* 18 (2018) 3.
8. U. Stimming, P. Bele, *J. Fuel Cells,* 17 (2017) 2.
9. T. Reddy, D. Linden, *Linden's Handbook of Batteries*, McGraw Hill, (2010) New York, United States.
10. N.H. Khalid, Y.M. Baba Ismail, A.A. Mohamad, *J. Power Sources, 176 (*2008) 393.
11. L. Ank T.S. Zhao, X.L. Zhou, X.H. Yan, C.Y. Jung, *J. Power Sources, 275 (* 2015) 831.
12. G. Li, M.A. Mezaal, R. Zhang, K. Zhang, L. Lei, *J. Fuel Cells*, 3 (2016) 395.

13. The Leclanché Cell, can be found under http://www.che.uc.edu/jensen/w.%20b.%20jensen/Museum%20Notes/24.%20The%20Leclanch%C3%A9%20Cell.pdf (2014).

14. E. Ahmed, *J. Adv. Res.*, 6 (2013) 105.

15. R. Sing, B. Veer, *Chemistry Select, 3 (*2018) 1309.

16. B. Crulhas, N. Ramos, C. Basso, V. Costa, G. Castro, V. Pedrosa, *Int. J. Electrochem. Sci.*, 9 (2014) 7596.

17. M. Chen, R. Wang, S. Cai, P. Mei, X. Yan, Y. Jiang, Y. Zhang, W. Xiao, H. Tang, *Int. J. Electrochem. Sci.*, 13 (2018) 2401.

18. M. Rosi, F. Iskandar, M. Abdullah, and Khairurrijal, *Int. J. Electrochem. Sci.*, 9 (2014) 4251.

19. Y.M.B. Ismail, H.Haliman, A.A. Mohamad, *Int. J. Electrochem. Sci.*, 7 (2012) 3555.

20. A. Mainar, E. Iruin, L. Colmenares, A. Kvasha, A. Kvasha, I. de Meatze, M. Bengoechea, O. Leonet, I. Boyano, Z. Zhang, J. Alberto Blazquez, *J.of Energy Storage, 15 (*2018) 304.

21. J-M. Kim, C. Kim, S. Yoo, J-H. Kim, J-H. Kim, J-M. Lim, S. Park, S-Y. Lee. *J. Mater. Chem.,3 (*2015) 10687.

22. D. McHugh, *Food and Agriculture Organization Of The United Nations, 288 (*1987) 1.

23. AnalogRead(), can be found under https://www.arduino.cc/reference/en/language/functions/analog-io/analogread/, (2017).

24. S. Wicelinski, M. Babiak, R. Runkles. *American National Standard For Portable Primary Cells and Batteries With Aqueous Electrolyte*, Part 1 (2001) 1.

25. Z. Rogulski, A. Czerwiński, *J. Solid State Electrochem., 7* (2001), 118.

26. J. O. Bockris, B. Conway, E. Yeager, R. White. *Comprehensive Treatise of Electrochemistry*, Plenum Press, (1981) New York, United States.

27. G. Li, M.A. Mezaal, R. Zhang, K. Zhang, L. Lei, *J. Fuel Cells,* 3 (2016) 395.

28. G. Vinal, *Primary Batteries*, Wiley, (1950) New York, United States.

29. G. Benson, J. Gluck, C. Kaufmann, *J. Electrochem. Soc., 90 (*1946) 441.

30. E. B. Sebok, R. L. Taylor, *Encyclopedia of Materials: Science and Technology*, Second Edition (2001) 902

31. G. Singh, M. Esmaeilpour, A. Ratner, *Fuel*, 246 (2019) 108

32. A. Celzard, J.F. Mareche, F. Payot, G. Furdin, *Carbon*, 40 (2002) 2801.

33. X. Zhang, Y. Cui, Z. Lv, M. Li, S. Ma, Z. Cui, Q. Kong, *Int. J. Electrochem. Sci.*, 6 (2011) 6063.

34. B. Marinho, M. Ghislandi, E. Tkalya, C. Koning, G. de With, *Powder Technology*, 221 (2012) 351

35. Energizer, Alkaline Manganese Dioxide Handbook and Application Manual, http://data.energizer.com/pdfs/alkaline_appman.pdf, (2018)

36. G.Ghiurcan, C-C. Liu, A. Webber, F. Freddrix, *J. Electrochem. Sci.*, 150 (2003) 922.

**Figure Captions**

Figure 1: A single-battery test setup

Figure 2: Schematic of data acquisition system

Figure 3: Gel Leclanché-graphite cell constructed in a 20 ml vial.

Figure 4: Closed gel Leclanché cell construction in an 11 ml glass chamber with ports for purging/filling and ports for current collection.

Figure 5: Gel Leclanché cell continuous discharge data across 10 kΩ resistor with *GR* as conductive additive. Average of three trials.

Figure 6: Gel Leclanché cell continuous discharge data across a 10 kΩ resistor with *AB* as the conductive additive, 0.5 g MnO₂, and variable zinc 0.25 g (🔷) 0.5 g (🟧) 0.75 g (🔺).

Figure 7: Gel Leclanché cell continuous discharge data across a 10 kΩ resistor with *GR* as the conductive additive, 0.5 g MnO₂, and variable zinc 0.25 g (🔷) 0.5 g (🟧) 0.75 g (🔺).

Figure 8. Energy density of cells made with variable amounts of zinc and different conductive additives (AB and GR).

Figure 9. Total milliamp hours of cells made with variable amounts of zinc and different conductive additives (AB and GR)

Figure 10. Coulombic efficiency as a % of theoretical coulombic capacity.

Figure 11. Fill and purge voltage of a closed gel Leclanché cell under 10 kΩ load with *AB* conductive additive. The cell was purged and refilled at 400000 seconds.